\documentclass[twocolumn,prb,aps,showpacs,amsmath,amssymb]{revtex4}

\usepackage{graphicx}
\usepackage{dcolumn}
\usepackage{bm}

\renewcommand{\vec}[1]{{\bf#1}}

\begin{document}

\title{Mass modification of itinerant carriers in RKKY oscillations induced
by finite range exchange interactions}

\author{Sergey Smirnov}
\affiliation{Institut f\"ur Theoretische Physik, Universit\"at Regensburg,
  D-93040 Regensburg, Germany}

\date{\today}

\begin{abstract}
We investigate the Ruderman-Kittel-Kasuya-Yosida oscillations of the itinerant
carrier spin density  in a system where those oscillations appear only due to
a finite distribution of a localized spin. The system represents a
half-infinite one-dimensional quantum wire with a magnetic impurity located at
its edge. In contrast to the conventional model of a point-like exchange
interaction the itinerant carrier spin density oscillations in this system
exist. We {\it analytically} demonstrate that when the radius of the exchange
interaction is less than the wave length of the itinerant carriers living on
the Fermi surface, the long range behavior of the oscillations is identical to
the one taking place in the zero radius limit of the same exchange interaction
but for an infinite one-dimensional quantum wire where, in comparison with
the original  half-infinite system, the mass of the itinerant carriers is
{\it strongly} modified by the exchange interaction radius. On the basis
of our analysis we make a suggestion on directionality of surface
Ruderman-Kittel-Kasuya-Yosida interaction shown in recent experiments: we
believe that in general the anisotropy of the Ruderman-Kittel-Kasuya-Yosida
interaction could result not only from the anisotropy of the {\it Fermi
surface} of itinerant carriers but also from the anisotropy of the {\it spin
carrying atomic orbitals} of magnetic impurities.
\end{abstract}


\maketitle

\section{Introduction}
Systems with interacting localized magnetic moments play an important role in
modern spintronics \cite{Zut_Fab_Sar}. Among those systems the ones where the
interaction between localized spins is due to the itinerant carriers are very
attractive for semiconductor spintronics because the magnetic properties of
the systems can be manipulated by the itinerant carrier density. The latter in
semiconductors can be easily varied using external gate voltage.

This, so-called itinerant magnetism, is provided by the exchange interaction
introduced by Ruderman and Kittel \cite {Rud_Kit} in the context of the
hyperfine interaction between a localized nuclear spin and conduction
electrons and later by Kasuya and Yosida \cite{Kasuya,Yosida} in the context
of the $s-d$ interaction between conduction electrons and magnetic ions of
transition metals.

The essence of the Ruderman-Kittel-Kasuya-Yosida (RKKY) itinerant magnetism is
that the exchange interaction between localized spins and paramagnetic
itinerant carrier system induces oscillations of the itinerant carrier spin
density. These oscillations propagating in an originally non-magnetic system
couple localized spins and create a magnetic structure. This type of
interaction becomes relevant when the direct spin-spin coupling is weak,
{\it i.e.}, in diluted systems. For example ferromagnetism in bulk diluted
magnetic semiconductors \cite{Dietl} is of the RKKY type. It was investigated
in Ga$_{1-x}$Mn$_x$As using high-resolution scanning tunneling microscope
\cite{Kitchen}.

The RKKY effect has also been widely studied in mesoscopic systems both
without spin-orbit interactions
\cite{Egger,Pershin,Utsumi,Rikitake,Tamura,Semiromi} and with spin-orbit
interactions \cite{Imamura,Simonin,Aono,Schulz}. In the latter case spin-orbit
interactions make the RKKY spin density oscillations anisotropic. Impact of
electron-electron interactions on the RKKY range function in one-dimensional
quantum wires (1D QW) was considered in Refs. \onlinecite{Egger,Schulz}. It
was shown that the range function is different from the one in non-interacting
1D QWs. It turns out that due to electron-electron interactions the range
function decays slower than in the non-interacting case.
\begin{figure}
\includegraphics[width=8.5 cm]{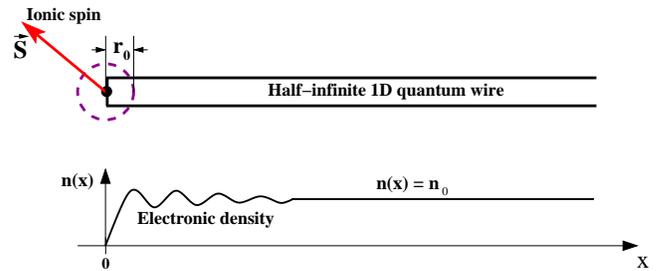}
\caption{\label{figure_1} (Color online) A half-infinite 1D QW with a magnetic
ion placed at its edge taken as a reference point. The ionic spin is $\vec{S}$
and $r_0$ characterizes its spatial distribution. The electronic density is
zero at the edge, $x=0$. Thus for $r_0=0$ one expects no RKKY oscillations in
the system. However, for $x>0$ the electronic density exhibits the Friedel
oscillations converging to the electronic density of the corresponding
infinite 1D QW, $n_0$. Therefore for $r_0\neq 0$ the exchange interaction
takes place and is responsible for the RKKY oscillations of the electron spin
density.}
\end{figure}

However, in spite of considerable amount of scientific research on the RKKY
interaction a little attention has been paid to the fact that the exchange
interaction responsible for the RKKY interaction is actually non-local in
space but has a finite range $r_0$ given by the spatial spin distribution
around the localization center. In theoretical models of the RKKY interaction
it is almost always assumed that $r_0=0$ and the exchange interaction has a
point-like character which is modeled by the Dirac delta-function. Of course,
in many cases it is a very good approximation when the wave function of a
magnetic impurity is well localized which means that $r_0$ is less than all
other relevant lengths present in the system. However, in semiconductors there
is one important length, namely the wave length of the itinerant carriers
living on the Fermi surface, $l_F$, which can be varied in a wide range. This
can be done, {\it e.g.}, by changing the carrier density with the help of a
gate voltage. Therefore, one can reach a situation where $l_F<r_0$.

What is more interesting is that even in the case when $l_F>r_0$ it can happen
that in a given system the RKKY spin density oscillations appear only due to a
finite value of $r_0$ and cannot appear if $r_0=0$. Such a situation takes
place, for example, when magnetic impurities are deposited on a surface of a
paramagnetic bulk sample. The wave function of the itinerant carriers in the
sample is equal to zero on its surface and therefore for a point-like model of
the exchange interaction, $r_0=0$, the RKKY spin density oscillations do not
appear and cannot couple the spins of the magnetic impurities on the
surface. Therefore, no magnetic ordering will take place. However, in reality
$r_0\neq 0$ and the exchange interaction will definitely produce the spin
density oscillations with the subsequent magnetic structure of the impurity
spins on the surface of the sample. Thus, in this system a finite value of
$r_0$ creates itinerant surface magnetism.

Some efforts to consider a finite range of the exchange interactions were made
in the research devoted to ferromagnetism in Heusler alloys
\cite{Darby,Smit,Darby_1,Malmstroem,Smit_1}. An essential drawback of the
models of the exchange interactions in those attempts was that in the limit
$r_0\rightarrow 0$ the models did not reproduce point-like exchange
interactions. However, it is desirable to reproduce the Dirac delta-function
in the limit $r_0\rightarrow 0$ because it is often used in many theoretical
models and therefore one would like to make a comparison with traditional
theories in the limit $r_0\rightarrow 0$.

A finite range exchange interaction which in the limit $r_0\rightarrow 0$
gives the traditional Dirac delta-function model has been used in
Ref. \onlinecite{Szalowski} to calculate the RKKY exchange integrals and in
Ref. \onlinecite{Smirnov} to study the RKKY spin density oscillations in a
mesoscopic ring. In particular, in Ref. \onlinecite{Szalowski} it has been
demonstrated that a finite value of $r_0$ removes an unphysical divergency of
the RKKY integral at zero distance while in Ref. \onlinecite{Smirnov} it has
been shown that any finite value of $r_0$ always becomes important if the number
of the electrons in the ring is large enough. However, the research in
Ref. \onlinecite{Smirnov} represented mainly a numerical experiment and, as
a result, did not clarify the physical role played by the localization radius
of the impurity spin. In addition, the RKKY spin density oscillations in that
study were also present in the limiting case $r_0\rightarrow 0$ and therefore
a finite value of $r_0$ did not play any key role in the formation of the RKKY
spin density oscillations.

In the present work we use the model of Ref. \onlinecite{Smirnov} for a finite range
exchange interaction in a system where the RKKY spin density oscillations
appear only if $r_0\neq 0$. The system represents a half-infinite 1D QW with a
magnetic impurity on its edge. Since the wave function of the itinerant
carriers (electrons to be definite) is zero on the edge of the wire, the RKKY
spin density oscillations exist only if $r_0$ is finite. To understand the
physical role played by the impurity spin localization radius in the RKKY spin
density oscillations we solve the problem {\it analytically} in the case
when $l_F>r_0$ and calculate the coordinate dependence of the RKKY
oscillations in this system. We demonstrate that at large distances this
dependence is identical to the one of the RKKY oscillations which one would
obtain in the limiting case $r_0=0$ of our model but for an infinite 1D QW
with electrons whose mass is {\it strongly} modified by the localization
radius of the impurity spin.

The paper is organized as follows. In Section \ref{FP} we mathematically
formulate the problem and solve it in Section \ref{DS} using the Feynman
diagram approach. In connection with recent experiments we make a suggestion
in Section \ref{SDSRKKYI} about directionality of surface RKKY
interaction. Conclusions are given in Section \ref{S}.

\section{Formulation of the problem}\label{FP}
As mentioned in the introduction the system represents (see
Fig. \ref{figure_1}) a half-infinite 1D QW with, {\it e.g.}, a magnetic ion on
its edge. It could be also another object, natural or artificial ({\it e.g.},
a quantum dot with odd electron number) with non-vanishing total spin
$\vec{S}$. The electrons in a half-infinite 1D QW have eigenenergies
\begin{equation}
\epsilon_{q_x}=\frac{q_x^2}{2m},
\label{en_sp}
\end{equation}
where $m$ is the effective mass of the conduction electrons and $q_x$ is not the
electron momentum because the translational invariance is broken. The quantum
number $q_x$ takes all real values except zero, $q_x\neq 0$. The eigenstates
$|q_x\sigma\rangle$, where $\sigma$ is the spin quantum number, in the coordinate
representation are
\begin{equation*}
\langle x\sigma'|q_x\sigma\rangle=
\begin{cases}
\delta_{\sigma'\sigma}\frac{1}{\sqrt{L}}\sin\bigl(\frac{1}{\hbar}q_xx\bigl)& x\geqslant 0\\
0& x<0
\end{cases},
\end{equation*}
where $L$ is the size of the system. To consider a half-infinite system we
take the limit $L\rightarrow\infty$ in the subsequent calculations. Since the
momentum operator is $\hat{p}_x=-i\hbar\partial/\partial x$, one can see that
$\langle x\sigma'|q_x\sigma\rangle$ is not the eigenfunction of $\hat{p}_x$ and
thus, as mentioned above, the quantum number $q_x$ is not the electron momentum.

The electronic density vanishes on the edge of the wire. Far from the edge
it has the value which it would have in an infinite 1D QW with the Fermi
momentum $q_F$ (in the infinite case it would be real momentum),
\begin{equation}
n_0=\frac{2q_F}{\pi\hbar}.
\label{den_bulk}
\end{equation}
In the region close to the edge the electronic density shows oscillations known
as the Friedel oscillations.

The half-infinite quantum wire can be modeled by the following potential:
\begin{equation}
u(x)=
\begin{cases}
0,&x>0,\\
\infty,&x\leqslant 0.
\end{cases}
\label{w_pot}
\end{equation}

The second quantized Hamiltonian of the half-infinite 1D QW without the magnetic
impurity thus reads in the coordinate basis as
\begin{equation}
\begin{split}
&\hat{H}_0=\sum_\sigma\int dx\:\hat{\psi}_\sigma^\dagger(x)\biggl[-\frac{\hbar^2}{2m}\frac{d^2}{dx^2}\biggl]\hat{\psi}_\sigma(x)+\\
&+\sum_\sigma\int dx\:u(x)\hat{\psi}_\sigma^\dagger(x)\hat{\psi}_\sigma(x),
\end{split}
\label{H_0}
\end{equation}
where $\hat{\psi}_\sigma^\dagger(x)$, $\hat{\psi}_\sigma(x)$ are the
electronic creation and annihilation field operators of the half-infinite 1D
QW.

When the magnetic impurity is deposited on the edge of the half-infinite 1D
QW, the electronic spin density interacts through the exchange interaction
with the impurity spin. The corresponding second quantized Hamiltonian is
\begin{equation}
 \hat{H}_{ex}=\sum_{\sigma,\sigma'}S^i\langle\sigma|\hat{\sigma}^i|\sigma'\rangle
\int dx\:J(x)\hat{\psi}_\sigma^\dagger(x)\hat{\psi}_{\sigma'}(x),
\label{H_ex}
\end{equation}
where $\hat{\vec{\sigma}}$ is the vector of the Pauli matrices and for $J(x)$
we use the model of Ref. \onlinecite{Smirnov},
\begin{equation}
J(x)=\frac{J}{r_0\sqrt{\pi}}\exp\biggl[-\biggl(\frac{x}{r_0}\biggl)^2\biggl].
\label{J_mod}
\end{equation}

The total Hamiltonian describing the RKKY electronic spin density oscillations
is given by the sum
\begin{equation}
\hat{H}=\hat{H}_0+\hat{H}_{ex}.
\label{H_tot}
\end{equation}

To get the RKKY oscillations and understand the physical role of the finite
impurity spin distribution $r_0$ on those oscillations one may consider
$\hat{H}_{ex}$ as a perturbation. In the next section we will implement this
perturbation theory on the language of the Feynman diagram expansion.
\begin{figure}
\includegraphics[width=8.5 cm]{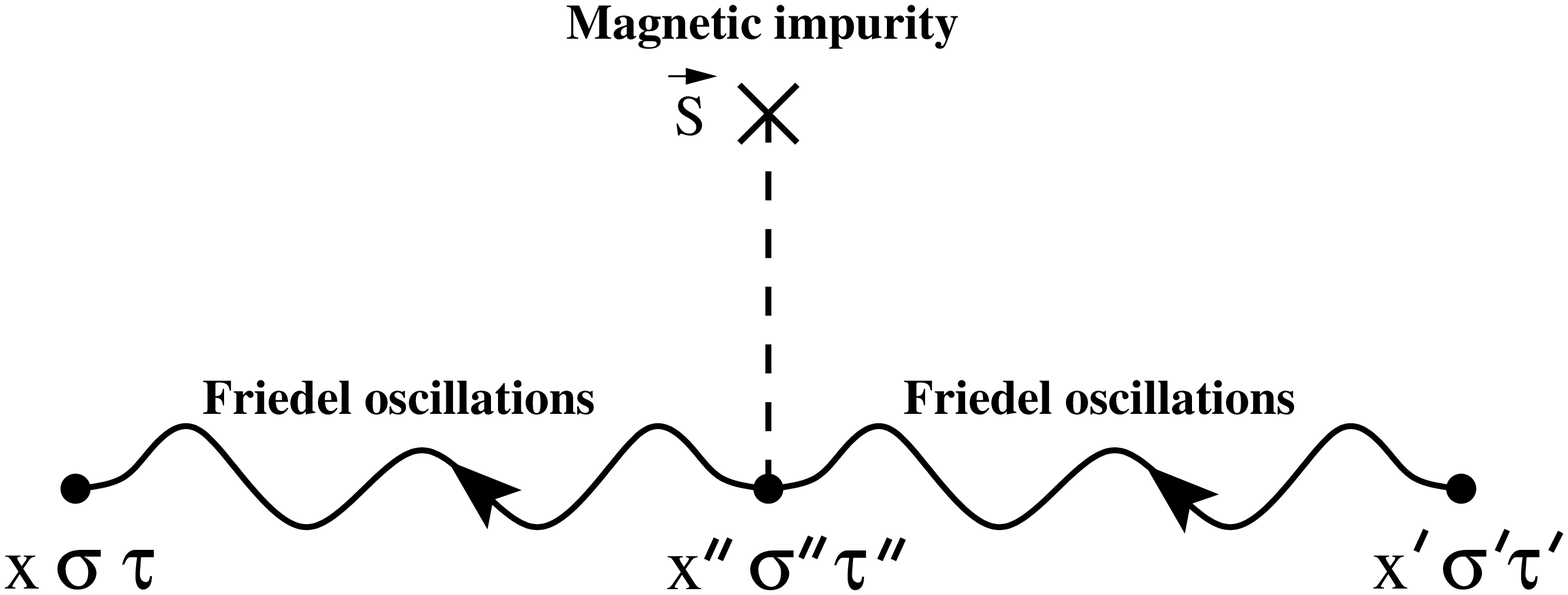}
\caption{\label{figure_2} (Color online) The first order Feynman diagram
  contributing to the electronic spin density oscillations. Unlike the
  traditional way to draw the propagators by straight lines the incoming and
  outgoing propagators are drawn using wavy lines to stress that they take
  into account the Friedel oscillations of the electronic density which are
  present in a half-infinite 1D QW without the magnetic impurity.}
\end{figure}

\section{Diagrammatic solution}\label{DS}
To obtain the electronic spin density we first have to find the imaginary-time
(or Matsubara) one-particle Green's function,
\begin{equation}
\mathcal{G}_{\sigma\sigma'}(x\tau|x'\tau')=\langle T\hat{\psi}_{\sigma}(x,\tau)\hat{\psi}_{\sigma'}^\dagger(x',\tau')\rangle,
\label{ITGF}
\end{equation}
where the angular brackets stand for the thermal average and the electronic
field operators are in the imaginary-time Heisenberg representation,
\begin{equation}
\begin{split}
&\hat{\psi}_{\sigma}(x,\tau)=e^{\tau(\hat{H}-\mu\hat{N})}\hat{\psi}_{\sigma}(x)e^{-\tau(\hat{H}-\mu\hat{N})},\\
&\hat{\psi}_{\sigma}^\dagger(x,\tau)=e^{\tau(\hat{H}-\mu\hat{N})}\hat{\psi}_{\sigma}^\dagger(x)e^{-\tau(\hat{H}-\mu\hat{N})},
\end{split}
\label{ITHR}
\end{equation}
where $\mu$ is the chemical potential.

The exact electronic spin density is obtained from the imaginary-time Green's
function as
\begin{equation}
\sigma^i_{exact}(x)=-\text{Tr}[\hat{\sigma}^i\mathcal{G}(x\tau|x\tau+0)],
\label{SD}
\end{equation}
where the trace is taken in the spin space.

The RKKY oscillations of the electronic spin density are contained in the
contribution to $\mathcal{G}_{\sigma\sigma'}(x\tau|x'\tau')$ of the first
order in the perturbation Hamiltonian $\hat{H}_{ex}$. The corresponding
Feynman diagram is shown in Fig. \ref{figure_2}. The wavy lines in this
diagram are the electronic propagators
$\mathcal{G}_{\sigma\sigma'}^{(0)}(x\tau|x'\tau')$ of the half-infinite 1D
QW without the magnetic impurity. They include the Friedel oscillations of the
electronic density and can be found using a trick suggested in
Ref. \onlinecite{LS}. The trick is based on the image charge technique
\cite{Landau_VIII}. The result is
\begin{equation}
\mathcal{G}_{\sigma\sigma'}^{(0)}(x\tau|x'\tau')=
\bar{\mathcal{G}}_{\sigma\sigma'}^{(0)}(x\tau|x'\tau')-
\bar{\mathcal{G}}_{\sigma\sigma'}^{(0)}(x\tau|-x'\tau')
\label{G_0a}
\end{equation}
for $x\geqslant 0$ and $x'\geqslant 0$,
\begin{equation}
\mathcal{G}_{\sigma\sigma'}^{(0)}(x\tau|x'\tau')=0
\label{G_0b}
\end{equation}
for $x<0$ or $x'<0$. In Eq. (\ref{G_0a})
$\bar{\mathcal{G}}_{\sigma\sigma'}^{(0)}(x\tau|x'\tau')$ is the imaginary-time
Green's function of the corresponding infinite 1D QW without the magnetic
impurity.

Using the rules \cite{AGD} for the analytic reading of Feynman diagrams we
obtain from the diagram shown in Fig. \ref{figure_2} the first order
contribution to $\mathcal{G}_{\sigma\sigma'}(x\tau|x'\tau')$, {\it i.e.},
$\mathcal{G}_{\sigma\sigma'}^{(1)}(x\tau|x'\tau')$ and calculate the
electronic spin density
\begin{equation}
\sigma^i(x)=-\text{Tr}[\hat{\sigma}^i\mathcal{G}^{(1)}(x\tau|x\tau+0)].
\label{SD_1}
\end{equation}
It is easily verified that the result is given by the following expression:
\begin{equation}
\begin{split}
&\sigma^i(x)=\frac{16mJS^i}{\pi^2\hbar^2}\times\\
&\times\int_0^\infty dq_x\int_0^\infty
dq'_x\sin\biggl(\frac{1}{\hbar}q_xx\biggl)\sin\biggl(\frac{1}{\hbar}q'_xx\biggl)\times\\
&\times Q_{q_xq'_x}\frac{n_{q'_x}-n_{q_x}}{q_x^{'2}-q_x^2},
\end{split}
\label{SD_any_T}
\end{equation}
where $x\geqslant 0$, $n_{q_x}$ is the Fermion occupation number,
\begin{equation}
n_{q_x}=\frac{1}{e^{\beta(\epsilon_{q_x}-\mu)}+1}
\label{FON}
\end{equation}
with $\beta=1/k_BT$ being the inverse temperature and the matrix $Q_{q_xq'_x}$
is
\begin{equation}
Q_{q_xq'_x}=\frac{1}{4}\biggl[e^{-\frac{r_0^2}{4\hbar^2}(q_x-q'_x)^2}-e^{-\frac{r_0^2}{4\hbar^2}(q_x+q'_x)^2}\biggl].
\label{Q_qxq'x}
\end{equation}
Therefore the dependence of the electronic spin density on the impurity spin
distribution comes through the matrix $Q_{q_xq'_x}$.

One can see from Eq. (\ref{Q_qxq'x}) that when $r_0=0$, the matrix
$Q_{q_xq'_x}$ vanishes and no RKKY spin density oscillations appear in the
system. This is in agreement with our earlier qualitative discussions.

Below we only consider the most interesting case of zero temperature
$T=0$. Using Eq. (\ref{Q_qxq'x}) we obtain from Eq. (\ref{SD_any_T}) in the
zero temperature limit the following expression for the electronic spin
density:
\begin{equation}
\begin{split}
&\sigma^i(x)=\frac{8mJS^i}{\pi^2\hbar^2}\times\\
&\times\int_0^1d\tilde{q}_x\int_1^\infty
  d\tilde{q}'_x\sin\biggl(\frac{q_Fx}{\hbar}\tilde{q}_x\biggl)
\sin\biggl(\frac{q_Fx}{\hbar}\tilde{q}'_x\biggl)\times\\
&\times\frac{e^{-\frac{r_0^2q_F^2}{4\hbar^2}(\tilde{q}_x-\tilde{q}'_x)^2}-e^{-\frac{r_0^2q_F^2}{4\hbar^2}(\tilde{q}_x+\tilde{q}'_x)^2}}
{\tilde{q}_x^{2}-\tilde{q}_x^{'2}},
\end{split}
\label{SDZT}
\end{equation}
where $\tilde{q}_x$ and $\tilde{q}'_x$ are dimensionless integration
variables.

The integral
\begin{equation}
\begin{split}
&I(x)=\\
&=\int_0^1d\tilde{q}_x\int_1^\infty
  d\tilde{q}'_x\sin\biggl(\frac{x}{l_F}\tilde{q}_x\biggl)
\sin\biggl(\frac{x}{l_F}\tilde{q}'_x\biggl)\times\\
&\times\frac{e^{-(\frac{r_0}{2l_F})^2(\tilde{q}_x-\tilde{q}'_x)^2}-e^{-(\frac{r_0}{2l_F})^2(\tilde{q}_x+\tilde{q}'_x)^2}}
{\tilde{q}_x^{2}-\tilde{q}_x^{'2}},
\end{split}
\label{I_int}
\end{equation}
where $l_F=\hbar/q_F$, can be calculated analytically in the case
$r_0/l_F\ll 1$. To do this we perform the expansion
\begin{equation}
e^{-(\frac{r_0}{2l_F})^2(\tilde{q}_x-\tilde{q}'_x)^2}-e^{-(\frac{r_0}{2l_F})^2(\tilde{q}_x+\tilde{q}'_x)^2}\approx
\biggl(\frac{r_0}{l_F}\biggl)^2\tilde{q}_x\tilde{q}'_x.
\label{Expansion}
\end{equation}
Let us denote through $J(x)$ the integral $I(x)$ approximated using this
expansion,
\begin{equation}
\begin{split}
&J(x)=\biggl(\frac{r_0}{l_F}\biggl)^2\times\\
&\times\int_0^1d\tilde{q}_x\int_1^\infty d\tilde{q}'_x
\sin\biggl(\frac{x}{l_F}\tilde{q}_x\biggl)\sin\biggl(\frac{x}{l_F}\tilde{q}'_x\biggl)\times\\
&\times\frac{\tilde{q}_x\tilde{q}'_x}{\tilde{q}_x^{2}-\tilde{q}_x^{'2}}.
\end{split}
\label{J_int}
\end{equation}
\begin{figure}
\includegraphics[width=8.5 cm]{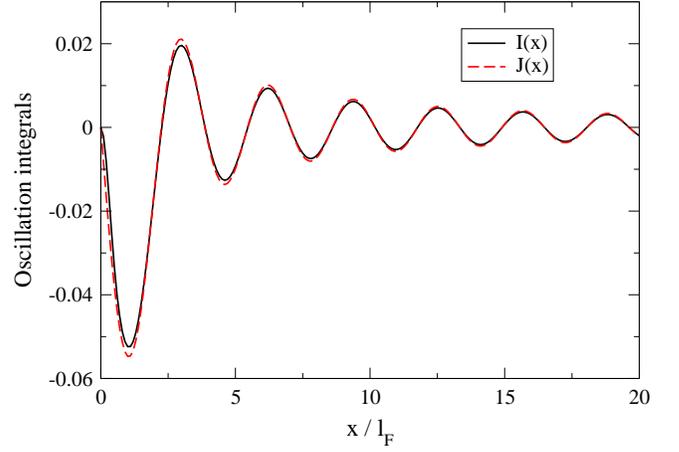}
\caption{\label{figure_3} (Color online) Comparison of the oscillation
  integrals $I(x)$ and $J(x)$ for $r_0/l_F=0.4$. The maximal deviation of the
  approximate integral $J(x)$ from the exact integral $I(x)$ takes place at
  short distances and is around 14\%.}
\end{figure}
\begin{figure}
\includegraphics[width=8.5 cm]{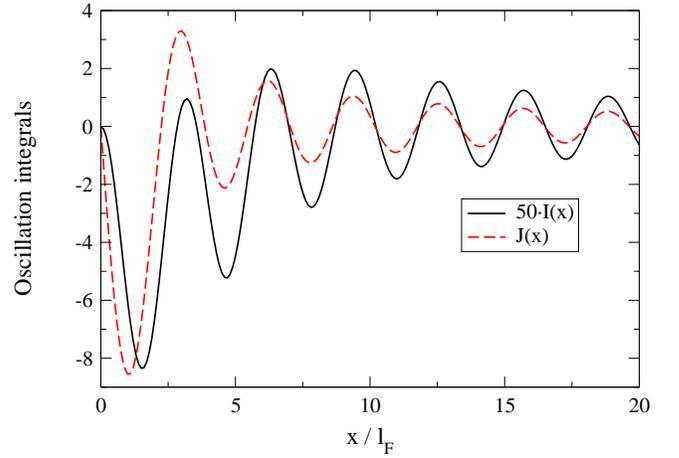}
\caption{\label{figure_4} (Color online) Comparison of the oscillation
  integrals $I(x)$ and $J(x)$ for $r_0/l_F=5$. The approximate integral $J(x)$
  greatly overestimates the exact integral $I(x)$. Moreover, the short range
  behavior of $J(x)$ is even incorrect qualitatively.}
\end{figure}
The analytical expression for the integral $J(x)$ is
\begin{equation}
J(x)=\frac{\pi}{16}\biggl(\frac{r_0}{l_F}\biggl)^2
\biggl[\frac{2\cos(2x/l_F)}{x/l_F}-\frac{\sin(2x/l_F)}{(x/l_F)^2}\biggl].
\label{J_int_anl}
\end{equation}
To demonstrate that the integral $J(x)$ is really a good approximation to the
exact integral $I(x)$ we calculate the integral $I(x)$ numerically and show
two situations in Figs. \ref{figure_3} and \ref{figure_4} for the two cases
$r_0/l_F<1$ and $r_0/l_F>1$, respectively. Even when $r_0/l_F=0.4$, {\it
  i.e.}, is not too much smaller than $1$, the integral $J(x)$ gives a good
precision and for smaller $r_0/l_F$ the difference between $J(x)$ and $I(x)$
decreases rapidly. On the other side, as one can see from Fig. \ref{figure_4},
in the case $r_0/l_F=5$ the integral $J(x)$ gives values much larger than the
exact integral $I(x)$ and at small distances it even predicts a wrong
qualitative coordinate dependence.

In the case when $r_0/l_F<1$ we therefore obtain the RKKY oscillations of the
electronic spin density,
\begin{equation}
\sigma^i(x)\!\!=\!\!JS^i\biggl(\frac{r_0}{l_F}\biggl)^2\frac{2m}{\pi\hbar^2}
\!\biggl[\frac{\cos(2x/l_F)}{2x/l_F}\!-\!\frac{\sin(2x/l_F)}{(2x/l_F)^2}\biggl].
\label{SD_ZTR0LTLF}
\end{equation}

In diluted magnetic systems one is interested in the long range behavior of
the RKKY oscillations. From Eq. (\ref{SD_ZTR0LTLF}) it follows that the long
range behavior of the RKKY oscillations is
\begin{equation}
\sigma^i(x)=JS^i\biggl(\frac{r_0}{l_F}\biggl)^2\frac{2m}{\pi\hbar^2}\frac{\cos(2x/l_F)}{2x/l_F}.
\label{SD_ZTR0LTLFLR}
\end{equation}

Comparison of the long range behavior of the RKKY spin density oscillations,
Eq. (\ref{SD_ZTR0LTLFLR}), with the one in a point-like Dirac's delta-function
model of the exchange interaction in an infinite 1D QW \cite{Yafet} allows us
to make the following conclusion. The RKKY oscillations induced by finite
range exchange interactions, $r_0\neq 0$, in a half-infinite 1D QW are
identical to the ones which would take place in the limit $r_0=0$ of our model
but in an infinite 1D QW with conduction electrons having a modified mass
\begin{equation}
m^\star=\biggl(\frac{r_0}{l_F}\biggl)^2m.
\label{RM}
\end{equation}
This modification is quite strong because in semiconductors the quantity
$r_0/l_F$ may be tuned in a wide range and it can be made much less than $1$.

Eq. (\ref{RM}) clarifies the physical role of the finite magnetic impurity
spin distribution in the formation of the RKKY spin density oscillations. One
can still use a point-like Dirac's delta-function model of the exchange
interaction between the magnetic impurity spin and conduction
electrons. However, the electron mass $m$ must be replaced by the modified
mass $m^\star$.

\section{A suggestion on directionality of surface RKKY interaction}\label{SDSRKKYI}
In recent experiments \cite{Zhou} surface RKKY interaction was studied using
cobalt adatoms on platinum (111). A strong directional dependence of the RKKY
spin density oscillations was observed. The anisotropic behavior was
attributed only to the anisotropy of the Fermi surface of platinum.

In connection with our analysis we would like to note that since the wave
function of the itinerant carriers in platinum is equal to zero on its
surface, the RKKY spin density oscillations in the experiments of
Ref. \onlinecite{Zhou} are definitely only due to a finite spin distribution
of cobalt adatoms. As we have shown above for a simple one-dimensional model,
the RKKY spin density oscillations existing only due to a finite value of
$r_0$ are created by electrons whose mass is strongly modified by the
radius of the exchange interaction $r_0$. Therefore, in addition to the Fermi
surface anisotropy the anisotropic behavior of the RKKY spin density
oscillations could also be produced by the fact that the itinerant carrier
mass for different directions suffers different modification.

It is not difficult to generalize our model, Eq. (\ref{J_mod}), to take into account  the
anisotropy of the spin carrying atomic orbitals of a magnetic impurity. The
simplest possibility for the case of magnetic adatoms on a two-dimensional
surface is
\begin{equation}
J(x,y)=\frac{J}{r_0^xr_0^y\pi}\exp\biggl[-\biggl(\frac{x}{r_0^x}\biggl)^2-\biggl(\frac{y}{r_0^y}\biggl)^2\biggl],
\label{J_mod_gen}
\end{equation}
where $r_0^x\neq r_0^y$. Other more adequate models taking into account the
real angular distribution of the magnetic impurity spin are also possible.

Since in general the symmetries of the Fermi surface of itinerant carries and
the spin carrying atomic orbitals of a magnetic impurity are different, it is
experimentally challenging to distinguish these two different kinds of
symmetries in the RKKY spin density oscillations. The experimental observation
of traces of the symmetry of the spin carrying atomic orbitals of a magnetic
impurity could prove that a finite spin distribution of the impurity spin
plays a significant role in the formation of the RKKY interaction.

\section{Summary}\label{S}
We have studied the Ruderman-Kittel-Kasuya-Yosida (RKKY) spin density
oscillations in a system where these oscillations appear only due to a finite
spin distribution of a magnetic impurity. It has been shown that the physical
role played by the finite spin distribution is to modify the mass of the
itinerant carriers in the system.

Making use of this result we have made a suggestion on the anisotropic
behavior of surface RKKY interaction observed in recent experiments which were
explained only from the point of view of the anisotropy of the Fermi surface
of itinerant carriers. We assume that the anisotropy of the spin carrying
atomic orbitals of a magnetic impurity could result in an anisotropic
itinerant carrier mass modification and thus it could also contribute to
an anisotropic behavior of surface RKKY interaction but with the impurity
atomic orbital symmetry which is in general different from the symmetry of the
Fermi surface of the itinerant carriers.

\begin{acknowledgments}
A discussion with Prof. Milena Grifoni and support from the DFG under the
program SFB 689 are acknowledged.
\end{acknowledgments}


\begin{thebibliography}{28}
\expandafter\ifx\csname natexlab\endcsname\relax\def\natexlab#1{#1}\fi
\expandafter\ifx\csname bibnamefont\endcsname\relax
  \def\bibnamefont#1{#1}\fi
\expandafter\ifx\csname bibfnamefont\endcsname\relax
  \def\bibfnamefont#1{#1}\fi
\expandafter\ifx\csname citenamefont\endcsname\relax
  \def\citenamefont#1{#1}\fi
\expandafter\ifx\csname url\endcsname\relax
  \def\url#1{\texttt{#1}}\fi
\expandafter\ifx\csname urlprefix\endcsname\relax\def\urlprefix{URL }\fi
\providecommand{\bibinfo}[2]{#2}
\providecommand{\eprint}[2][]{\url{#2}}

\bibitem[{\citenamefont{Zutic et~al.}(2004)\citenamefont{Zutic, Fabian, and
  {Das Sarma}}}]{Zut_Fab_Sar}
\bibinfo{author}{\bibfnamefont{I.}~\bibnamefont{Zutic}},
  \bibinfo{author}{\bibfnamefont{J.}~\bibnamefont{Fabian}}, \bibnamefont{and}
  \bibinfo{author}{\bibfnamefont{S.}~\bibnamefont{{Das Sarma}}},
  \bibinfo{journal}{Rev. Mod. Phys.} \textbf{\bibinfo{volume}{76}},
  \bibinfo{pages}{323} (\bibinfo{year}{2004}).

\bibitem[{\citenamefont{Ruderman and Kittel}(1954)}]{Rud_Kit}
\bibinfo{author}{\bibfnamefont{M.~A.} \bibnamefont{Ruderman}} \bibnamefont{and}
  \bibinfo{author}{\bibfnamefont{C.}~\bibnamefont{Kittel}},
  \bibinfo{journal}{Phys. Rev.} \textbf{\bibinfo{volume}{96}},
  \bibinfo{pages}{99} (\bibinfo{year}{1954}).

\bibitem[{\citenamefont{Kasuya}(1956)}]{Kasuya}
\bibinfo{author}{\bibfnamefont{T.}~\bibnamefont{Kasuya}},
  \bibinfo{journal}{Progr.\ Theor.\ Phys.} \textbf{\bibinfo{volume}{16}},
  \bibinfo{pages}{45} (\bibinfo{year}{1956}).

\bibitem[{\citenamefont{Yosida}(1957)}]{Yosida}
\bibinfo{author}{\bibfnamefont{K.}~\bibnamefont{Yosida}},
  \bibinfo{journal}{Phys.\ Rev.} \textbf{\bibinfo{volume}{106}},
  \bibinfo{pages}{893} (\bibinfo{year}{1957}).

\bibitem[{\citenamefont{Dietl et~al.}(2000)\citenamefont{Dietl, Ohno,
  Matsukura, Cibert, and Ferrand}}]{Dietl}
\bibinfo{author}{\bibfnamefont{T.}~\bibnamefont{Dietl}},
  \bibinfo{author}{\bibfnamefont{H.}~\bibnamefont{Ohno}},
  \bibinfo{author}{\bibfnamefont{F.}~\bibnamefont{Matsukura}},
  \bibinfo{author}{\bibfnamefont{J.}~\bibnamefont{Cibert}}, \bibnamefont{and}
  \bibinfo{author}{\bibfnamefont{D.}~\bibnamefont{Ferrand}},
  \bibinfo{journal}{Science} \textbf{\bibinfo{volume}{287}},
  \bibinfo{pages}{1019} (\bibinfo{year}{2000}).

\bibitem[{\citenamefont{Kitchen et~al.}(2006)\citenamefont{Kitchen,
  Richardella, Tang, Flatte, and Yazdani}}]{Kitchen}
\bibinfo{author}{\bibfnamefont{D.}~\bibnamefont{Kitchen}},
  \bibinfo{author}{\bibfnamefont{A.}~\bibnamefont{Richardella}},
  \bibinfo{author}{\bibfnamefont{J.~M.} \bibnamefont{Tang}},
  \bibinfo{author}{\bibfnamefont{M.~E.} \bibnamefont{Flatte}},
  \bibnamefont{and} \bibinfo{author}{\bibfnamefont{A.}~\bibnamefont{Yazdani}},
  \bibinfo{journal}{Nature} \textbf{\bibinfo{volume}{442}},
  \bibinfo{pages}{436} (\bibinfo{year}{2006}).

\bibitem[{\citenamefont{Egger and Schoeller}(1996)}]{Egger}
\bibinfo{author}{\bibfnamefont{R.}~\bibnamefont{Egger}} \bibnamefont{and}
  \bibinfo{author}{\bibfnamefont{H.}~\bibnamefont{Schoeller}},
  \bibinfo{journal}{Phys.\ Rev.\ B} \textbf{\bibinfo{volume}{54}},
  \bibinfo{pages}{16337} (\bibinfo{year}{1996}).

\bibitem[{\citenamefont{Pershin et~al.}(2003)\citenamefont{Pershin, Vagner, and
  Wyder}}]{Pershin}
\bibinfo{author}{\bibfnamefont{Y.~V.} \bibnamefont{Pershin}},
  \bibinfo{author}{\bibfnamefont{I.~D.} \bibnamefont{Vagner}},
  \bibnamefont{and} \bibinfo{author}{\bibfnamefont{P.}~\bibnamefont{Wyder}},
  \bibinfo{journal}{J.\ Phys.:\ Condens.\ Matter}
  \textbf{\bibinfo{volume}{15}}, \bibinfo{pages}{997} (\bibinfo{year}{2003}).

\bibitem[{\citenamefont{Utsumi et~al.}(2004)\citenamefont{Utsumi, Martinek,
  Bruno, and Imamura}}]{Utsumi}
\bibinfo{author}{\bibfnamefont{Y.}~\bibnamefont{Utsumi}},
  \bibinfo{author}{\bibfnamefont{J.}~\bibnamefont{Martinek}},
  \bibinfo{author}{\bibfnamefont{P.}~\bibnamefont{Bruno}}, \bibnamefont{and}
  \bibinfo{author}{\bibfnamefont{H.}~\bibnamefont{Imamura}},
  \bibinfo{journal}{Phys.\ Rev.\ B} \textbf{\bibinfo{volume}{69}},
  \bibinfo{pages}{155320} (\bibinfo{year}{2004}).

\bibitem[{\citenamefont{Rikitake and Imamura}(2005)}]{Rikitake}
\bibinfo{author}{\bibfnamefont{Y.}~\bibnamefont{Rikitake}} \bibnamefont{and}
  \bibinfo{author}{\bibfnamefont{H.}~\bibnamefont{Imamura}},
  \bibinfo{journal}{Phys.\ Rev.\ B} \textbf{\bibinfo{volume}{72}},
  \bibinfo{pages}{033308} (\bibinfo{year}{2005}).

\bibitem[{\citenamefont{Tamura and Glazman}(2005)}]{Tamura}
\bibinfo{author}{\bibfnamefont{H.}~\bibnamefont{Tamura}} \bibnamefont{and}
  \bibinfo{author}{\bibfnamefont{L.}~\bibnamefont{Glazman}},
  \bibinfo{journal}{Phys.\ Rev.\ B} \textbf{\bibinfo{volume}{72}},
  \bibinfo{pages}{121308(R)} (\bibinfo{year}{2005}).

\bibitem[{\citenamefont{Semiromi and Ebrahimi}(2006)}]{Semiromi}
\bibinfo{author}{\bibfnamefont{E.~H.} \bibnamefont{Semiromi}} \bibnamefont{and}
  \bibinfo{author}{\bibfnamefont{F.}~\bibnamefont{Ebrahimi}},
  \bibinfo{journal}{Phys.\ Rev.\ B} \textbf{\bibinfo{volume}{73}},
  \bibinfo{pages}{195418} (\bibinfo{year}{2006}).

\bibitem[{\citenamefont{Imamura et~al.}(2004)\citenamefont{Imamura, Bruno, and
  Utsumi}}]{Imamura}
\bibinfo{author}{\bibfnamefont{H.}~\bibnamefont{Imamura}},
  \bibinfo{author}{\bibfnamefont{P.}~\bibnamefont{Bruno}}, \bibnamefont{and}
  \bibinfo{author}{\bibfnamefont{Y.}~\bibnamefont{Utsumi}},
  \bibinfo{journal}{Phys.\ Rev.\ B} \textbf{\bibinfo{volume}{69}},
  \bibinfo{pages}{121303(R)} (\bibinfo{year}{2004}).

\bibitem[{\citenamefont{Simonin}(2006)}]{Simonin}
\bibinfo{author}{\bibfnamefont{J.}~\bibnamefont{Simonin}},
  \bibinfo{journal}{Phys.\ Rev.\ Lett.} \textbf{\bibinfo{volume}{97}},
  \bibinfo{pages}{266804} (\bibinfo{year}{2006}).

\bibitem[{\citenamefont{Aono}(2007)}]{Aono}
\bibinfo{author}{\bibfnamefont{T.}~\bibnamefont{Aono}},
  \bibinfo{journal}{Phys.\ Rev.\ B} \textbf{\bibinfo{volume}{76}},
  \bibinfo{pages}{073304} (\bibinfo{year}{2007}).

\bibitem[{\citenamefont{Schulz et~al.}(2009)\citenamefont{Schulz, {De Martino},
  Ingenhoven, and Egger}}]{Schulz}
\bibinfo{author}{\bibfnamefont{A.}~\bibnamefont{Schulz}},
  \bibinfo{author}{\bibfnamefont{A.}~\bibnamefont{{De Martino}}},
  \bibinfo{author}{\bibfnamefont{P.}~\bibnamefont{Ingenhoven}},
  \bibnamefont{and} \bibinfo{author}{\bibfnamefont{R.}~\bibnamefont{Egger}},
  \bibinfo{journal}{Phys.\ Rev.\ B} \textbf{\bibinfo{volume}{79}},
  \bibinfo{pages}{205432} (\bibinfo{year}{2009}).

\bibitem[{\citenamefont{Darby}(1977{\natexlab{a}})}]{Darby}
\bibinfo{author}{\bibfnamefont{M.~I.} \bibnamefont{Darby}},
  \bibinfo{journal}{J.\ Phys. F: Metal Phys.} \textbf{\bibinfo{volume}{7}},
  \bibinfo{pages}{L69} (\bibinfo{year}{1977}{\natexlab{a}}).

\bibitem[{\citenamefont{Smit}(1977)}]{Smit}
\bibinfo{author}{\bibfnamefont{J.}~\bibnamefont{Smit}}, \bibinfo{journal}{J.\
  Phys. F: Metal Phys.} \textbf{\bibinfo{volume}{7}}, \bibinfo{pages}{L189}
  (\bibinfo{year}{1977}).

\bibitem[{\citenamefont{Darby}(1977{\natexlab{b}})}]{Darby_1}
\bibinfo{author}{\bibfnamefont{M.~I.} \bibnamefont{Darby}},
  \bibinfo{journal}{J.\ Phys. F: Metal Phys.} \textbf{\bibinfo{volume}{7}},
  \bibinfo{pages}{L191} (\bibinfo{year}{1977}{\natexlab{b}}).

\bibitem[{\citenamefont{Malmstr{\"o}m and Geldart}(1978)}]{Malmstroem}
\bibinfo{author}{\bibfnamefont{G.}~\bibnamefont{Malmstr{\"o}m}}
  \bibnamefont{and} \bibinfo{author}{\bibfnamefont{D.~J.~W.}
  \bibnamefont{Geldart}}, \bibinfo{journal}{J.\ Phys. F: Metal Phys.}
  \textbf{\bibinfo{volume}{8}}, \bibinfo{pages}{L17} (\bibinfo{year}{1978}).

\bibitem[{\citenamefont{Smit}(1978)}]{Smit_1}
\bibinfo{author}{\bibfnamefont{J.}~\bibnamefont{Smit}}, \bibinfo{journal}{J.\
  Phys. F: Metal Phys.} \textbf{\bibinfo{volume}{8}}, \bibinfo{pages}{2139}
  (\bibinfo{year}{1978}).

\bibitem[{\citenamefont{Szalowski and Balcerzak}(2008)}]{Szalowski}
\bibinfo{author}{\bibfnamefont{K.}~\bibnamefont{Szalowski}} \bibnamefont{and}
  \bibinfo{author}{\bibfnamefont{T.}~\bibnamefont{Balcerzak}},
  \bibinfo{journal}{Phys.\ Rev.\ B} \textbf{\bibinfo{volume}{78}},
  \bibinfo{pages}{024419} (\bibinfo{year}{2008}).

\bibitem[{\citenamefont{Smirnov}(2009)}]{Smirnov}
\bibinfo{author}{\bibfnamefont{S.}~\bibnamefont{Smirnov}},
  \bibinfo{journal}{Phys.\ Rev.\ B} \textbf{\bibinfo{volume}{79}},
  \bibinfo{pages}{134403} (\bibinfo{year}{2009}).

\bibitem[{\citenamefont{Levitov and Shitov}(2003)}]{LS}
\bibinfo{author}{\bibfnamefont{L.~S.} \bibnamefont{Levitov}} \bibnamefont{and}
  \bibinfo{author}{\bibfnamefont{A.~V.} \bibnamefont{Shitov}},
  \emph{\bibinfo{title}{Green's Functions. Problems and Solutions}}
  (\bibinfo{publisher}{Fizmatlit, Moscow}, \bibinfo{year}{2003}),
  \bibinfo{edition}{2nd} ed., \bibinfo{note}{in Russian}.

\bibitem[{\citenamefont{Landau and Lifshitz}(1984)}]{Landau_VIII}
\bibinfo{author}{\bibfnamefont{L.~D.} \bibnamefont{Landau}} \bibnamefont{and}
  \bibinfo{author}{\bibfnamefont{E.~M.} \bibnamefont{Lifshitz}},
  \emph{\bibinfo{title}{Electrodynamics of Continuous Media}},
  vol.~\bibinfo{volume}{8} (\bibinfo{publisher}{Butterworth-Heinemann},
  \bibinfo{year}{1984}), \bibinfo{edition}{2nd} ed.

\bibitem[{\citenamefont{Abrikosov et~al.}(1963)\citenamefont{Abrikosov, Gorkov,
  and Dzyaloshinski}}]{AGD}
\bibinfo{author}{\bibfnamefont{A.~A.} \bibnamefont{Abrikosov}},
  \bibinfo{author}{\bibfnamefont{L.~P.} \bibnamefont{Gorkov}},
  \bibnamefont{and} \bibinfo{author}{\bibfnamefont{I.~E.}
  \bibnamefont{Dzyaloshinski}}, \emph{\bibinfo{title}{Methods of Quantum Field
  Theory in Statistical Physics}} (\bibinfo{publisher}{Dover, New York},
  \bibinfo{year}{1963}).

\bibitem[{\citenamefont{Yafet}(1987)}]{Yafet}
\bibinfo{author}{\bibfnamefont{Y.}~\bibnamefont{Yafet}},
  \bibinfo{journal}{Phys.\ Rev.\ B} \textbf{\bibinfo{volume}{36}},
  \bibinfo{pages}{3948} (\bibinfo{year}{1987}).

\bibitem[{\citenamefont{Zhou et~al.}(2010)\citenamefont{Zhou, Wiebe, Lounis,
  Vedmedenko, Meier, Bl{\"u}gel, Dederichs, and Wiesendanger}}]{Zhou}
\bibinfo{author}{\bibfnamefont{L.}~\bibnamefont{Zhou}},
  \bibinfo{author}{\bibfnamefont{J.}~\bibnamefont{Wiebe}},
  \bibinfo{author}{\bibfnamefont{S.}~\bibnamefont{Lounis}},
  \bibinfo{author}{\bibfnamefont{E.}~\bibnamefont{Vedmedenko}},
  \bibinfo{author}{\bibfnamefont{F.}~\bibnamefont{Meier}},
  \bibinfo{author}{\bibfnamefont{S.}~\bibnamefont{Bl{\"u}gel}},
  \bibinfo{author}{\bibfnamefont{P.~H.} \bibnamefont{Dederichs}},
  \bibnamefont{and}
  \bibinfo{author}{\bibfnamefont{R.}~\bibnamefont{Wiesendanger}},
  \bibinfo{journal}{Nat.\ Phys.} \textbf{\bibinfo{volume}{6}},
  \bibinfo{pages}{187} (\bibinfo{year}{2010}).

\end{thebibliography}
\end{document}